\documentclass[12pt]{iopart}
\usepackage{graphicx}
\usepackage{epstopdf}
\usepackage{hyperref}
\usepackage[usenames]{color}
\expandafter\let\csname equation*\endcsname\relax
\expandafter\let\csname endequation*\endcsname\relax
\usepackage{amssymb,amsmath}
\usepackage[square,comma,compress,numbers]{natbib}
\newcommand{\eq}[1]{Eq.~(\ref{#1})}
\newcommand{\fig}[1]{Fig.~\ref{#1}}
\newcommand{\be}[1]{\begin{equation}\label{#1}}
\newcommand{\ee}{\end{equation}}

\usepackage{float}

\renewcommand{\url}[1]{\href{#1}{Link}}

\newcommand{\gj}[6]{ \begin{pmatrix}
 #1 & #2 & #3 \\
 #4 & #5 & #6 
 \end{pmatrix}}

\begin{document}
\title{Pulse dependence of  prevalent pathways in xenon driven by an x-ray  free-electron-laser pulse}
\author{Henry I. B. Banks}
\address{Department of Physics and Astronomy, University College London, Gower Street, London WC1E 6BT, United Kingdom}
\author{Antonis Hadjipittas}
\address{Department of Physics and Astronomy, University College London, Gower Street, London WC1E 6BT, United Kingdom}
\author{Agapi Emmanouilidou}
\address{Department of Physics and Astronomy, University College London, Gower Street, London WC1E 6BT, United Kingdom}

\begin{abstract}
We study the interaction of xenon with an 850 eV photon energy FEL pulse. We do so by  employing a Monte-Carlo technique. We compute the single-photon  ionisation cross sections and Auger rates, used in the Monte-Carlo technique, by adopting  to atoms a  formalism we previously developed for diatomic molecules. We determine the yields of the ion states  of driven xenon and compare with previously obtained experimental results.  To better understand the yields obtained, we identify  the prevalent  pathways leading to the formation of each final ion state of xenon. We gain further insight into the high yields of highly-charged ion states by comparing the yields and dominant pathways of these ion states of  xenon when  driven by different FEL pulses that have the same energy. We show that higher-charged ion states have higher yields when xenon is driven by longer-duration   pulses due to Auger cascades taking place between subsequent single-photon ionisations. 
 \end{abstract}
\date{\today}

\maketitle
 
\section{Introduction}

The development of x-ray free-electron lasers (XFELs) \cite{FELhistory,Bostedt2013,Emma2010} constitutes a new tool for high-resolution probing of atoms and molecules \cite{Marangos2011,Ullrich2012} and bio-molecular imaging \cite{Schlichting2012,Neutze2000,Redecke227,Santra2009}. The x-ray energy of these photons results in inner-shell electrons being more likely to ionise than valence ones. Inner-shell ionisation results in the creation of `hollow' states. Such states are of particular interest in molecules, due to the sensitivity of these states to their chemical environment \cite{Cederbaum1986,Cederbaum1987}. A `hollow' state can decay quickly via an Auger process. That is, an electron from a higher energy sub-shell falls down to fill in a core hole and in doing so releases enough energy to ionise another electron. The new state, resulting from an Auger transition, may also be a `hollow' state itself, thus, triggering an Auger cascade.

FEL interactions with xenon have been the subject of many studies in recent years \cite{Sorokin2007, Rudek2012, Makris2009, Lambropoulos2011, Toyota2017, Son2012}. Namely, time-of-flight experiments have measured the ion yields produced when xenon interacts with high-intensity FEL pulses at both low (extreme ultraviolet) \cite{Sorokin2007} and high (x-ray) \cite{Rudek2012} photon energies. For low photon energies, using rate equations, insights into the resulting highly-charged ion states have been obtained by accounting, in addition to single-photon, for multi-photon processes \cite{Makris2009, Lambropoulos2011}. The cross sections for these multi-photon processes were obtained using scaling laws \cite{Madsen1999}. For x-ray photon energies, Monte-Carlo methods have been employed to gain insight into the dynamics leading to the formation of ion states \cite{Son2012,Jurek2016}. Moreover, the interaction of xenon clusters with XFELs has been addressed and the ion yields obtained both experimentally and theoretically \cite{Thomas2012}. 

In this work, we gain further insight into the formation of the final ion states of xenon driven by an XFEL pulse by identifing the prevalent pathways for each ion state. A pathway refers to the sequence of states accessed in time due to the interaction of xenon with the pulse. Each state is described by its electronic configuration. In our previous studies of argon interacting with 260 eV and 315 eV photon energy FEL pulses, we obtained the ion yields and prevalent pathways by employing rate equations \cite{Wallis2014,Wallis2015}.
Here, we do so by employing a Monte-Carlo technique \cite{Robert2005}. The latter is chosen over rate equations due to the significantly larger number of states that are energetically accessed when xenon interacts with an 850 eV pulse compared to argon interacting with a 315 eV pulse. For argon roughly 200 states are energetically accessed, while for xenon this number  increases to roughly $2\times10^{4}$. This implies that $2\times10^{4}$ rate equations must be solved to obtain the xenon ion yields.  However, to identify the  pathways of xenon, the number of rate equations to be solved  is computationally prohibitive. To address this issue, we employ a much faster Monte-Carlo technique which, starting from an initial state, accesses only one state at a time, which is determined in a stochastic manner \cite{Jurek2016}. 
 
To implement the Monte-Carlo technique for xenon driven by an 850 eV FEL pulse, we  compute the single-photon ionisation cross sections and Auger decay rates for all energetically allowed transitions. We do so by adapting to atoms the formalism that we have previously developed for diatomic molecules \cite{Banks2017}. The formalism  developed in the current work  is general and applies to complex multi-electron  atoms   with orbital wave functions that can be accurately described when accounting for several  angular momentum quantum numbers $l$. In contrast, the previous formalism we employed for computing single-photon ionisation cross sections  and Auger rates \cite{Wallis2014,Wallis2015} is only applicable to atoms with orbital  wave functions  described by one quantum number $l$, as was the case for argon. 

Employing the single-photon ionisation cross sections and Auger rates, thus computed, in the Monte-Carlo technique, we obtain the ion yields and prevalent pathways in the formation of the ion states of xenon driven by an 850 eV pulse. We find these pathways to be more complex compared to the ones involved in argon driven by a 315 eV FEL pulse \cite{Wallis2014,Wallis2015}. In the latter work, we have found that even-charged ion states of argon were more populated than odd-charged ones. For argon interacting with an FEL pulse at 315 eV, creation of one inner-shell vacancy is followed by only one Auger decay. As a result, the prevalent pathways involve mostly sequences of the pair of an inner-shell single-photon ionisation plus an Auger decay.
However, in this study, an 850 eV photon allows up to three electrons in the $3d$ sub-shell of xenon to ionise by single-photon absorption. Importantly, the creation of each $3d$ inner-shell hole can be followed by a cascade of up to five Auger transitions \cite{Jonauskas2003}. These Auger transitions involve electrons from shells with principal quantum number $n=4$ and $n=5$ filling in holes in shells with $n=3$ and $n=4$. Hence, the resulting dynamics for driven xenon is more complex compared to driven argon. This complex dynamics is imprinted in features of the ion yields and dominant pathways. We obtain the ion yields of xenon for parameters of FEL pulses where experimental results are available \cite{Thomas2012}. To understand the ion yields obtained, we  identify the prevalent pathways for each final ion state of xenon. In addition, keeping the energy of the FEL pulse constant, we address how these dominant pathways change for varying parameters of the FEL pulse. 


\section{Method}
\label{Method}


We employ a Monte-Carlo technique to obtain the yields of the final ion states and the pathways leading to the formation of each state. In our computations, neutral xenon and each ion state of xenon is defined by the electronic configuration $1s^{a}2s^{b}2p^{c}3s^{d}3p^{e}3d^{f}4s^{g}4p^{h}4d^{k}5s^{l}5p^{m}$. Each of the $np$ sub-shells involves orbitals $np_x,np_y,np_z$, while each of the $nd$ sub-shells involves orbitals $nd_{xy},nd_{yz},nd_{xz},$ $nd_{x^2-y^2},nd_{z^2} $. The maximum occupancy of each orbital is two. In what follows, we obtain the single-photon ionisation cross sections and Auger rates in terms of orbitals. We then sum over the relevant orbitals to obtain the respective rates between sub-shells. These latter rates are the ones employed in the Monte-Carlo technique. 

\subsection{Single-photon ionisation}
\label{Method:Xsecs}

We compute the single-photon ionisation cross section for an electron to transition from an initial bound orbital ${\phi_a}$ to a final continuum orbital ${\phi_{\epsilon,l,m}}$, with well-defined  quantum number $l$ and magnetic quantum number $m$, as follows \cite{ModernQM}:

 \begin{equation}
 \label{photoXsec}
\sigma_{a\rightarrow \epsilon,l,m} = \frac{4}{3}\alpha\pi^2 \omega N_a \sum_{M=-1,0,1} {|D_{a \rightarrow \epsilon,l,m}^M|}^{2}.
\end{equation}

\noindent We denote by ${\alpha}$ the fine structure constant, by ${N_a}$ the occupation number of the bound orbital $a$, by ${\omega}$ the photon energy, and by $M$ the polarisation of the photon. In the length gauge, the matrix element $D_{a \rightarrow \epsilon,l,m}^M$ is given by

\begin{equation}
\label{cross}
{D_{a \rightarrow \epsilon,l,m}^M = \int\phi_a({\bf r})\phi^*_{\epsilon,l,m}({\bf r})\sqrt{\frac{4\pi}{3}}rY_{1M}(\theta,\phi)d{\bf r}.}
\end{equation}

For computational efficiency, we split the integral in Eq.\eq{cross} into a radial and an angular component, since the latter integral can be performed analytically. To do so, we express the orbital $\phi_{a}$ in terms of radial wavefunctions, $P_{a,l',m'}(r)$, and spherical harmonics, $Y_{l',m'}(\theta,\phi)$ as follows

\begin{equation}
\label{eq:SCE}
\begin{aligned}
\phi_{a}({\bf r})=\sum_{l',m'}P_{a,l',m'}(r)Y_{l',m'}(\theta,\phi)/r. 
\end{aligned}
\end{equation}

\noindent We compute the bound orbital wavefunctions using Molpro \cite{molpro} with the AQZP \cite{Martins2013} basis set, which results in an accurate description of the bound orbitals of xenon. Employing this basis set, each bound orbital is expressed as a combination of several $l',m'$ terms. In contrast, in our studies for driven argon, each orbital had well defined $l,m$ numbers. The reason was that we employed the 6-311G basis set which was sufficient to adequately describe the bound orbitals of argon. Furthermore, we express the continuum orbital $\phi_{\epsilon,l,m}$ as follows

\begin{equation}
\label{eq:SCEcont}
\begin{aligned}
\phi_{\epsilon,l,m}({\bf r})=P_{\epsilon,l}(r)Y_{l,m}(\theta,\phi)/r. 
\end{aligned}
\end{equation}

\noindent We compute the continuum wavefunctions, $\phi_{\epsilon,l,m}$, using the Numerov technique \cite{Numerov1924} with a Hartree-Fock-Slater (HFS) potential obtained from the Herman-Skillman code \cite{Herman1963,Pauli2001}. We also employed this technique to obtain continuum orbitals in previous studies of argon driven by an FEL pulse \cite{Wallis2014}.

Substituting in Eq.\eq{cross} the expressions for $\phi_a({\bf r})$ and $\phi_{\epsilon,l,m}({\bf r})$ given by Eq.\eq{eq:SCE} and Eq.\eq{eq:SCEcont}, respectively, we obtain

\begin{equation}
\label{eq:pi_wig}
\begin{aligned}
D_{a \rightarrow \epsilon,l,m}^M = & \sqrt{\frac{4\pi}{3}}\sum_{l',m'}\int^\infty_0 dr P_{a,l',m'}(r) r P^*_{\epsilon,l}(r)
\\ \times & \int d\Omega Y_{l',m'}(\theta,\phi)Y^*_{l,m}(\theta,\phi)Y_{1M}(\theta,\phi).
\end{aligned}
\end{equation}

\noindent The angular part can be expressed analytically in terms of Wigner-3j symbols \cite{AMinQM}, leading to

\begin{equation}
\label{eq:pi_wig2}
\begin{aligned}
 D_{a \rightarrow \epsilon,l,m}^M&= \sum_{l',m'}{(-1)}^{m} \sqrt{(2l+1)(2l'+1)}
 \\&\times
 \begin{pmatrix} 
{l'} & {l} & {1} \\ 
{0} & {0} & {0}
\end{pmatrix} \begin{pmatrix}
{l'} & {l} & {1} \\ 
{-m'} & {m} & {M}
\end{pmatrix} 
\\
&\times\int^\infty_0 dr P_{a,l',m'}(r) r P^*_{\epsilon,l}(r).&&
\end{aligned}
\end{equation}

\begin{table}
\caption{Single-photon ionisation cross sections for different sub-shells of xenon in the ground state interacting with an 850 eV pulse.}
\label{tab:PhotoComp1}
\begin{center}
\begin{tabular}{lll}
\hline\noalign{\smallskip}
initial sub-shell & Ref. \cite{Yeh1985}[cm\textsuperscript{2}] & This Work[cm\textsuperscript{2}]\\ 
\noalign{\smallskip}\hline\noalign{\smallskip}
$3d^{10}$ & 2.05e-18 & 1.61e-18\\ 
$4s^2$ & 3.77e-20 & 3.10e-20\\ 
$4p^6$ & 1.27-19 & 9.60e-20\\
$4d^{10}$ & 2.34-19 & 1.79e-19\\
$5s^2$ & 6.04-21 & 3.68e-21\\
$5p^6$ & 1.42-20 & 8.56e-21\\
\noalign{\smallskip}\hline
\end{tabular}
\end{center}
\end{table}

\noindent For the transitions considered in our computations, only the energy of the final continuum orbital is of relevance. Hence, we sum over $l$ and $m$ in Eq.\eq{eq:pi_wig2} to obtain

\begin{equation}
\label{photoXsec}
\sigma_{a\rightarrow \epsilon} = \frac{4}{3}\alpha\pi^2 \omega N_a \sum_{l,m} \sum_{M=-1,0,1} {|D_{a \rightarrow \epsilon,l,m}^M|}^{2}.
\end{equation}

\noindent Finally, to obtain the single-photon ionisation cross section from a sub-shell $nl$, we sum $\sigma_{a\rightarrow \epsilon}$ over the orbitals $a$ in this sub-shell.

In table \ref{tab:PhotoComp1}, we compare our results for the single-photon ionisation cross sections of xenon with previous theoretical work \cite{Yeh1985}. We find a reasonable agreement, with the results in Ref. \cite{Yeh1985}, which employs a HFS method for the computation of both bound and continuum orbitals. We employ a Hartree-Fock method to compute the bound states and a HFS method to compute the continuum orbitals. We find cross sections that are roughly 80\% of the ones obtained in Ref. \cite{Yeh1985}. The agreement is not as good for the cross sections involving valence orbitals. The basis set we employ describes valence orbitals less accurately than inner-shell orbitals. However, for 850 eV photon energy, single-photon ionisation cross sections of valence orbitals are very small compared to the ones of inner-shell orbitals rendering irrelevant the contribution of the former to our computations. 

\subsection{Auger decay}
\label{Method:Auger}

In general, the Auger rate is given by \cite{FermiGR}
 \begin{equation}
 \begin{aligned}
\Gamma=\overline{\sum}2\pi |\mathcal{M}|^2\equiv\overline{\sum}2\pi |\langle\Psi_{fin}|H_I|\Psi_{init}\rangle|^2,&&
\end{aligned}
\label{eq:GeneralAuger}
\end{equation}

\noindent where $\overline{\sum}$ denotes a summation over the final states and an average over the initial states. $|\Psi\rangle$ is the wavefunction of all electrons in the atomic state. We assume that the Auger transition is a two-electron process, with $H_I$ being the electron-electron Coulomb repulsion term. We adapt to atoms  the derivation used in our previous work for molecules \cite{Banks2017}. We then find that the Auger rate involving two valence orbitals $a$ and $b$, an inner-shell orbital $c$, and a continuum orbital $\epsilon$ with quantum numbers $l,m$ is given by

\begin{equation}
\begin{aligned}
&\mathcal{M}=\delta_{S',S}\delta_{M',M}\sum_{\substack{ l_c,m_c,k\\l_a,m_a,l_b,m_b}}\sum_{q=-k}^{k}\int{dr_1}\int{dr_2}\\
&(-1)^{m+m_c+q}\sqrt{(2l+1)(2l_c+1)(2l_b+1)(2l_a+1)}\\
&\left[ P_{\epsilon,l}(r_1) P_{c,l_c,m_c}(r_2)\frac{r^k_<}{r^{k+1}_>}P_{b,l_b,m_b}(r_1)P_{a,l_a,m_a}(r_2) 
\vphantom{\gj{l_c}{k}{l_a}{0}{0}{0}\gj{l_c}{k}{l_a}{-m_c}{q}{m_a}}
\right.\\
&\left.
\gj{l}{k}{l_b}{0}{0}{0}\gj{l}{k}{l_b}{-m}{-q}{m_b}
\gj{l_c}{k}{l_a}{0}{0}{0}\gj{l_c}{k}{l_a}{-m_c}{q}{m_a}\right.\\
&+(-1)^S\left. P_{\epsilon,l}(r_1) P_{c,l_c,m_c}(r_2)\frac{r^k_<}{r^{k+1}_>}P_{a,l_a,m_a}(r_1)P_{b,l_b,m_b}(r_2)
\vphantom{\gj{l_c}{k}{l_a}{0}{0}{0}\gj{l_c}{k}{l_a}{-m_c}{q}{m_a}}
\right.\\
&\left.
\gj{l}{k}{l_a}{0}{0}{0}\gj{l}{k}{l_a}{-m}{-q}{m_a}
\gj{l_c}{k}{l_b}{0}{0}{0}\gj{l_c}{k}{l_b}{-m_c}{q}{m_b}\right],
\end{aligned}
\label{SM_S4intsSCE}
\end{equation}

\noindent where $r_<=\min(r_1,r_2)$, $r_>=\max(r_1,r_2)$. Moreover, $k$ and $q$ are the angular and magnetic quantum numbers, respectively, of the spherical harmonics involved in the multipole expansion of the Coulomb interaction, i.e. in the $1/r_{12}$ term. The total Auger rate is given by

\begin{equation}
\begin{aligned}
\Gamma_{b,a\rightarrow c,\epsilon}&=
\sum_{\substack{S,M_S\\S',M'_S}}\pi N_{ab}N_h\sum_{l,m}|\mathcal{M}|^2,
\end{aligned}
\label{TotalSM_S}
\end{equation}

\noindent where $N_h$ is the number of core holes in orbital c and $N_{ab}$ is the normalisation factor given by

\begin{equation}
\begin{aligned}
N_{ab}=&\frac{N_aN_b}{2\times2} && \mathrm{valence\, electrons\, in\, different\, orbitals}\\
=&\frac{N_a(N_a-1)}{2\times2\times1} &&\mathrm{valence\, electrons\, in\, the\ same\, orbital}.
\end{aligned}
\label{molN_12}
\end{equation}
\noindent $N_a$ and $N_b$ denote the occupation numbers of orbitals $a$ and $b$, respectively. Here, $S$ denotes the total spin of the two valence electrons, while $S'$ denotes the total spin of the core electron plus the electron in the continuum. $M_S$ and $M_S'$ denote their projections. To obtain the total Auger rate, $\Gamma_{s,t\rightarrow u,\epsilon}$ between $s,t,u$ sub-shells, we add the Auger rates $\Gamma_{b,a\rightarrow c,\epsilon}$ over the $a$ and $b$ orbitals in the respective $s,t$ sub-shells. We do not add over the $c$ orbitals in sub-shell $u$, since we average over the initial states.

\begin{table}[h]
\caption{Comparison of Auger rates, expressed in units of $10^{-2}$ a.u. The valence electron refers to an electron in any of the sub-shells $4s,4p,4d,5s$ or $5p$.}
\label{tab:AugerComp2}
\begin{center}
\begin{tabular}{llll}
 \hline\noalign{\smallskip}
$\Gamma_{s,t\rightarrow u,\epsilon}$ & Ref. \cite{McGuire1972} & Ref. \cite{Son2012} & This Work \\ 
\noalign{\smallskip}\hline\noalign{\smallskip}
$u=3s$ & & & \\
\hspace{.4cm}$s= val,t= val$ & $2.06$ & $1.85$ & $1.71$\\
\hspace{.4cm}$s= 3p,t= val$ & $27.8$ & $47.6$ & $45.6$\\
\hspace{.4cm}$s= 3d,t= val$ & $7.59$ & $8.98$ & $8.61$\\
$u=3p$ hole & & & \\
\hspace{.4cm}$s= val,t= val$ & $2.18$ & $2.10$ & $2.03$\\
\hspace{.4cm}$s= 3d,t= val$ & $17.0$ & $20.6$& $19.2$\\
$u=3d$ hole & & & \\ 
\hspace{.4cm}$s= val,t= val$ & $2.49$ & $2.26$ & $2.31$\\ 
\noalign{\smallskip}\hline
\end{tabular}
\end{center}
\end{table}

\begin{table}[h]
\caption{Comparison of Auger rates, expressed in units of $10^{-4}$ a.u. Each of the values given below corresponds to a sum of Auger rates, $\Gamma_{s,t\rightarrow u,\epsilon}$, over the relevant $s$ and $t$ sub-shells in the respective $n$ shells.}
\label{tab:AugerComp3}
\begin{center}
\begin{tabular}{llll}
 \hline\noalign{\smallskip}
 $\sum_{s,t}\Gamma_{s,t\rightarrow u,\epsilon}$ & Ref. \cite{McGuire1972} & This Work \\
\noalign{\smallskip}\hline\noalign{\smallskip}
$u=2s$ & & &\\
\hspace{.4cm}$n=3,3$	& 491 	& 465\\
\hspace{.4cm}$n=4,4$		& 10.7 	& 8.40\\
\hspace{.4cm}$n=5,5$ 	& 0.078 	& 0.043\\
\hspace{.4cm}$n=3,4$ 	& 143 	& 123.3\\
\hspace{.4cm}$n=3,5$ 	& 12.4 	& 7.79\\
\hspace{.4cm}$n=4,5$ 	& 1.78 	& 1.15\\
$u=2p$ & & &\\
\hspace{.4cm}$n=3,3$	& 671	& 716\\
\hspace{.4cm}$n=4,4$	& 15.3	& 12.6\\
\hspace{.4cm}$n=5,5$ 	& 0.079 	& 0.043\\
\hspace{.4cm}$n=3,4$ 	& 201 	& 189\\
\hspace{.4cm}$n=3,5$ 	& 14.6 	& 8.49\\
\hspace{.4cm}$n=4,5$ 	& 3.99 	& 1.22\\
\noalign{\smallskip}\hline
\end{tabular}
\end{center}
\end{table}


In tables \ref{tab:AugerComp2} and \ref{tab:AugerComp3}, we compare our results for the Auger rates to previous calculations \cite{McGuire1972, Son2012}. The Auger rates in Ref. \cite{McGuire1972} are computed using semi-empirical methods and in Ref. \cite{Son2012} using a non-relativistic HFS method. In table \ref{tab:AugerComp2}, we find a close agreement with the results obtained in Ref. \cite{Son2012}. This is expected, since the techniques employed in our work are similar to the techniques used in Ref. \cite{Son2012}. In table \ref{tab:AugerComp3}, the agreement with the results in Ref. \cite{McGuire1972} is reasonable given that the techniques employed in Ref. \cite{McGuire1972} are semi-empirical and thus less accurate than the techniques employed in the current study.

\subsection{Shake-off}
\label{Method:ShakeOff}

Following single-photon ionisation, if the photo-electron escapes with high energy, the sudden change in the potential energy can result in the ionisation of one more electron. According to the sudden approximation \cite{Aberg1969, Carlson1973} the probability for an electron to be shaken off from a $u$ sub-shell is given by 

\begin{equation}
\label{shakeUp}
P^S_{u}\approx 1-\left[\left| \int\phi^*_{a}({\bf r})\phi_{a}'({\bf r})d{\bf r} \right|^2\right]^{N_u},
\end{equation}

\noindent where $\phi_a$ and $\phi_a'$ are the orbital eigenstates of the Hamiltonian before and after single-photon ionisation takes place, with $a$ an orbital of the $u$ sub-shell and $N_u$ the occupancy of the sub-shell. Note that in Eq.\eq{shakeUp}, we neglect shake-up processes involving transitions of an electron to a bound orbital that differs from $a$.


\subsection{Monte-Carlo technique}
\label{Method:Monte-Carlo}

This method involves starting from the ground state of xenon and propagating forward in time with a time step, $\delta t=0.01$ fs. We have checked that the Monte-Carlo technique converges for the chosen time step. At each time step, we consider all energetically allowed transitions from the current state $i$ to a state $j$. We then compute the rates associated with these allowed transitions, as follows

\begin{equation}
\begin{aligned}
w_{i\rightarrow j}(t)&=\sigma_{i\rightarrow j}J(t)\qquad &&\mathrm{photo\,\, ionisation}\\
&=\sigma_{i\rightarrow j}P^{S}J(t)\qquad &&\mathrm{photo\,\, ionisation\, +\, shake\,\, off}\\
&=\Gamma_{i\rightarrow j}\qquad &&\mathrm{Auger\,\,decay}.
\end{aligned}
\end{equation}

\noindent We do not include fluorescence in our calculations, since the fluorescence rates are found to be much lower than the Auger rates \cite{Son2012}. Assuming that each of these processes follows an exponential decay law \cite{Jurek2016}, we assign a time to each allowed transition as follows

\begin{equation}
t_{i\rightarrow j}(t)=-\ln(\chi)/w_{i\rightarrow j}(t),
\end{equation}

\noindent where $\chi$ is a random number between zero and one. If the smallest time $t_{i\rightarrow j}$ associated with these transitions is less than the propagation time step, we register the transition associated with the smallest time. Next,  we repeat the process starting from time $t+t_{i\rightarrow j}$ with xenon in state $j$. Following this process, we propagate forward until reaching asymptotic times. The above described process corresponds to one realisation of the Monte-Carlo method. We implement this method $10^5$ times, which we find to be a sufficient number of realisations to achieve convergence.

\section{Results}

\begin{figure}[h!] 
\resizebox{\textwidth}{!}{
 \includegraphics{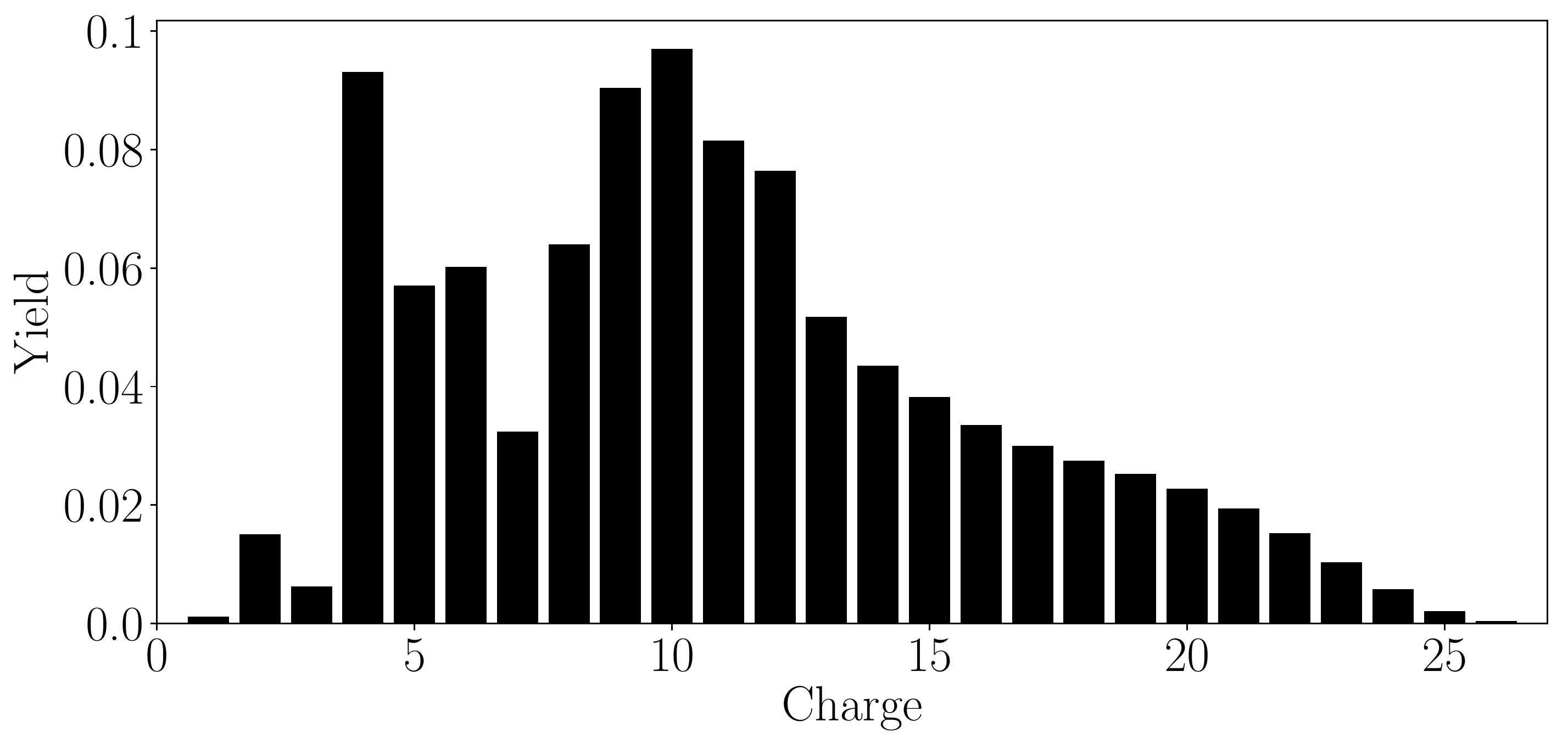}
 }
\caption{Final ion yields of xenon interacting with an 850 eV FEL pulse of peak intensity $5.6\times10^{16}$ $Wcm^{-2}$and duration 150 fs. The spatial distribution of the laser pulse is taken into account.}
\label{fig:IonYieldsComparison}
\end{figure}

We first obtain the final ion yields for xenon interacting with an 850 eV FEL pulse. To compare with the experimental results obtained in Ref. \cite{Thomas2012}, we use the same parameters for the FEL pulse as in Ref. \cite{Thomas2012}. That is, we consider a pulse of peak intensity $I_0$ equal to $5.6\times10^{16}$ $Wcm^{-2}$ and of full-width half-maximum $\tau$ equal to 150 fs. Accounting for the gaussian profile of the pulse in space, the resulting photon flux is given by 

\begin{flalign}
J(x,y,t) = I_0\exp{\left(-4ln2\left[(x/\rho_x)^2+(y/\rho_y)^2+(t/\tau)^2\right]\right)}/\omega,
\label{Jt1}
\end{flalign} 

\noindent with $\rho_x=2.2$ $\mu$m and $\rho_y=1.2$ $\mu$m being equal to the values used in the experiments in Refs. \cite{Thomas2012, Buth2012}. We compute the ion yields for the intensities corresponding to $(10^3)^2$ points on a square grid with dimensions $(10\,\mathrm{mm})^2$ and average over the grid area. 

Our results for the final ion yields are shown in \fig{fig:IonYieldsComparison}. The yields of the ion states $\mathrm{Xe}^{q+}$ with $q=1,2,3$ are very small. In contrast, the yields of the ion states with $q=4,5,6$ as well as with $q=\mathrm{8-13}$ are significant. We also find that the highest-charged state reaches up to $q=26$, in accord with the single-photon ionisation processes occurring up to charge $q=25$. These results are in agreement with the experimental results in Ref. \cite{Thomas2012}. 

\begin{figure}[h!] 
\resizebox{\textwidth}{!}{
 \includegraphics{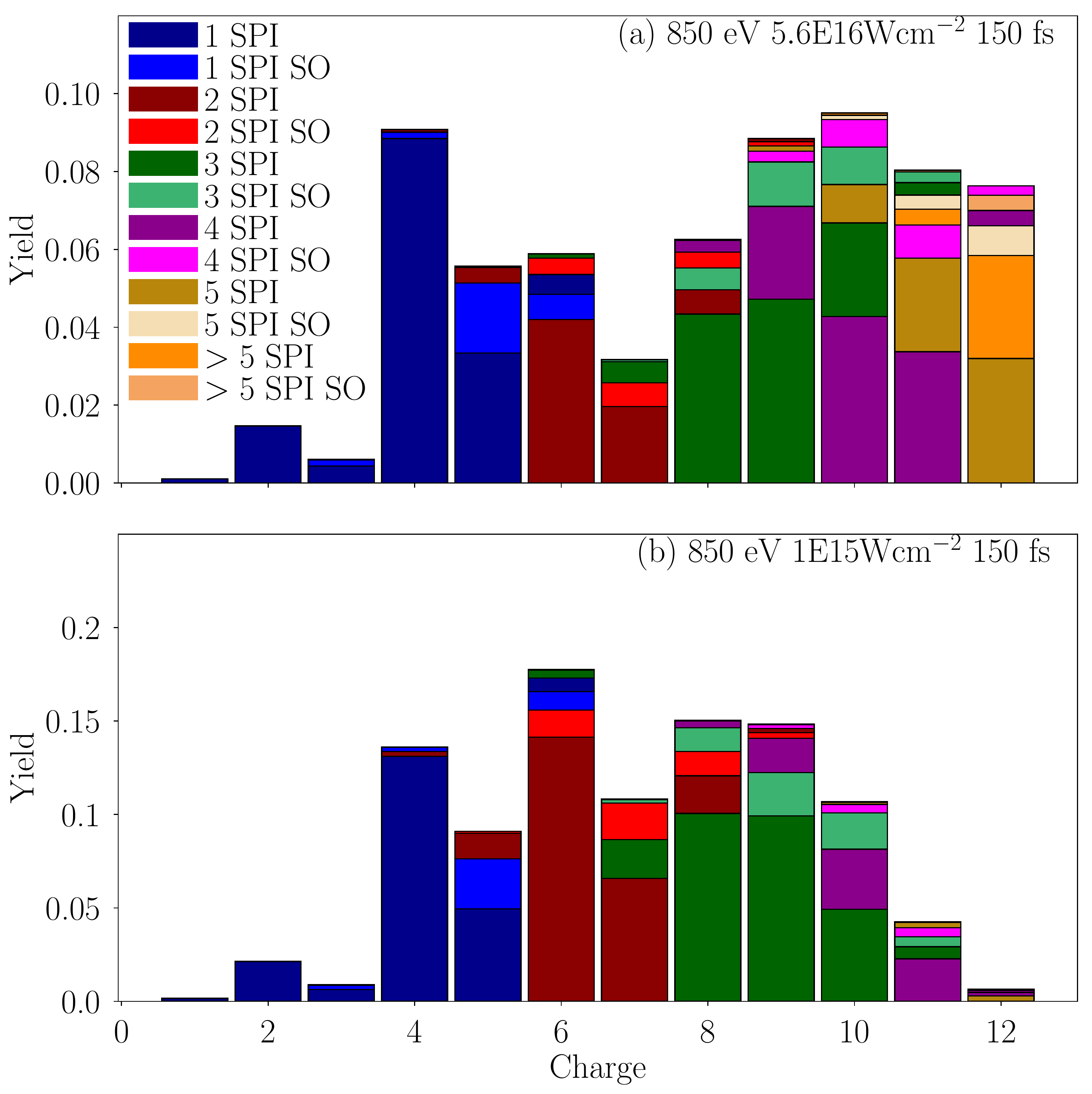}
 }
\caption{Final ion yields of xenon interacting with an 850 eV FEL pulse of duration 150 fs. For each ion state, we show the contribution of pathways separated according to the number of single-photon ionisation transitions that take place. (a) An FEL pulse of peak intensity $5.6\times10^{16}$ $Wcm^{-2}$ is considered and the results account for the spatial distribution of the pulse; (b) an FEL pulse of peak intensity $10^{15}$ $Wcm^{-2}$ is considered without accounting for the spatial distribution of the pulse. SPI stands for single-photon ionization, while SO stands for shake-off.}
\label{fig:2PathwaysUnordered}
\end{figure}

Next, we  identify the reason for the high yields of the ion states with $q$ around 10 for xenon driven by an 850 eV FEL pulse of peak intensity $5.6\times10^{16}$ $Wcm^{-2}$ and duration 150 fs. To do so, we compare the ion yields and pathways of xenon in Fig.  \ref{fig:2PathwaysUnordered}(a) with those  in  Fig. \ref{fig:2PathwaysUnordered}(b) when xenon is driven by an FEL pulse of smaller peak intensity $10^{15}$ $Wcm^{-2}$. In Figs. \ref{fig:2PathwaysUnordered}(a) and \ref{fig:2PathwaysUnordered}(b), in addition to the ion yields,  we show, for each ion state, the contribution of pathways that differ in the number  of single-photon absorptions that take place. As expected, we find that the pulse with higher peak intensity, results in a higher number of single-photon ionisation processes. This is clearly seen for the case of charges $q=10$ and $12$. The ion state with $q=10$ is mostly populated by pathways involving four single-photon ionisation processes, for the higher-intensity pulse, versus three, for the lower-intensity one. In addition, the ion state with $q=12$ has a significant yield only for the higher-intensity pulse, due to the large number, equal to five, of single-photon ionisations taking place. Hence, the pulse of the higher peak intensity has higher ion yields for the high-charged ion states due to the larger number of single-photon absorptions that occur. 

\begin{figure*}[h!] 
\resizebox{\textwidth}{!}{%
 \includegraphics{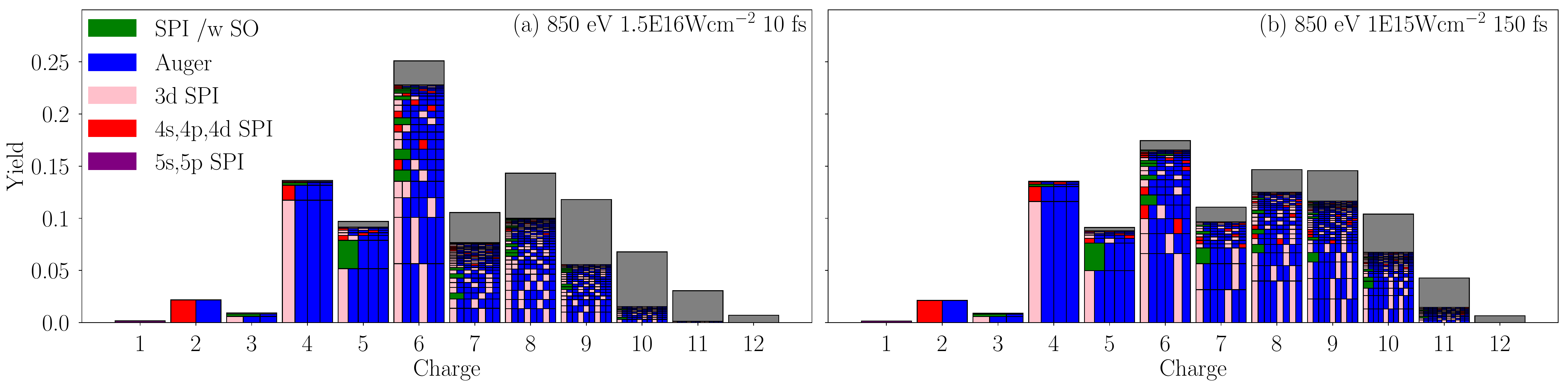}
 }
\caption{Final ion yields of  xenon interacting with 850 eV FEL pulses. For each ion state we show the prevalent  pathways. (a) An FEL pulse of duration 10 fs and peak intensity $1.5\times10^{16}$ $Wcm^{-2}$ is considered; (b) an FEL pulse of duration 150 fs and peak intensity $10^{15}$ $Wcm^{-2}$ is considered. Pathways with yield $<1\times10^{-3}$ are not distinguishable and are shown in grey.}
\label{PathwaysComparison} 
\end{figure*}

\begin{figure}[h!] 
\resizebox{\textwidth}{!}{
 \includegraphics{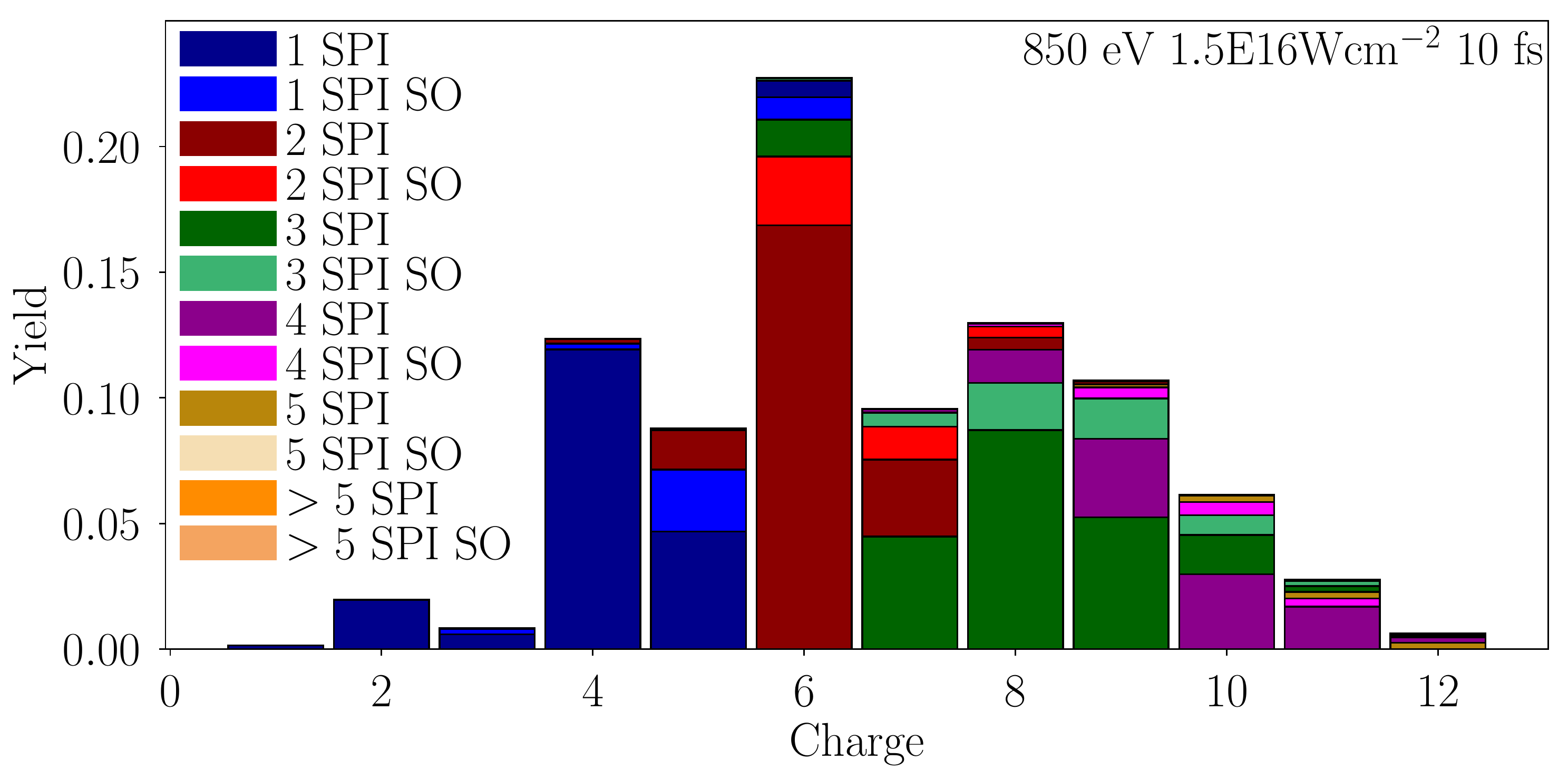}
 }
\caption{As for Fig. \ref{fig:2PathwaysUnordered}(b) but with an 850 eV FEL pulse of duration 10 fs and peak intensity $1.5\times10^{16}$ $Wcm^{-2}$.}
\label{fig:singlePathwayUnordered10fs}
\end{figure}

However, a large number of single-photon absorptions is not the only reason to obtain high yields for higher-charged ion states of xenon. To demonstrate this, we obtain the prevalent pathways for each ion state of xenon when driven by FEL pulses of the same energy. For instance, in Figs. \ref{PathwaysComparison}(a) and \ref{PathwaysComparison}(b), we compare the pathways prevalent in xenon driven by a short-duration, 10 fs, FEL pulse of peak intensity $1.5\times10^{16}$ $Wcm^{-2}$ versus a long duration, 150 fs, FEL pulse of peak intensity $10^{15}$ $Wcm^{-2}$. The results in Fig. \ref{PathwaysComparison} are obtained without accounting for the spatial dependence of the FEL pulses. We find that the yields of the ion states with charge $q=9,10$ are larger for the longer-duration and lower-intensity FEL pulse considered in Fig. \ref{PathwaysComparison}(b). This cannot be attributed to a larger number of single-photon absorptions occurring in the case of the longer pulse. Indeed, comparing Fig. \ref{fig:2PathwaysUnordered}(b) (same pulse as in Fig. \ref{PathwaysComparison}(b)) with Fig. \ref{fig:singlePathwayUnordered10fs} (same pulse as in Fig. \ref{PathwaysComparison}(a)), we find that a larger number of single-photon absorptions take place for the shorter-duration and higher-intensity FEL pulse compared to the longer-duration and lower-intensity pulse.

A close inspection of Figs. \ref{PathwaysComparison}(a) and (b) reveals that the ion states with $q=8,9,10$ have prevalent pathways that differ in the number of Auger transitions taking place between two subsequent photo-ionisation processes. Namely, for the shorter-duration pulse in Fig. \ref{PathwaysComparison}(a), the higher values of the photon flux reached at earlier times, compared to the longer-duration pulse in Fig. \ref{PathwaysComparison}(b), result in subsequent single-photon absorptions taking place in a shorter time interval. As a result, for the shorter-duration pulse in Fig. \ref{PathwaysComparison}(a), a smaller number of Auger decays take place between subsequent single-photon ionisation events. In turn, this results in the shell with principal quantum number $n=4$ being more depleted for the shorter-duration pulse by the time the next photon absorption takes place. In contrast, for the longer-duration pulse, with more Auger transitions taking place between single-photon absorptions, $4d$ electrons fill in $3d$ core holes, as for the shorter-duration pulse, but in addition, $5s$ and $5p$ electrons fill in the $4d$ holes. Hence, by the time a second photon absorption takes place, there are more electrons in the $n=4$ shell compared to the shorter-duration pulse. Since, the Auger rate is proportional to the occupation number, the rate of $n=4$ electrons filling in a $3d$ hole is higher for the longer pulse. Thus, the larger number of Auger transitions in between subsequent single photon absorptions is the main reason for the higher ion yields of states with $q=9,10$ for the longer duration pulse in Fig. \ref{PathwaysComparison}(b).

\section{Conclusions}

In this work, we have employed a Monte-Carlo technique that accounts for single-photon ionisations, Auger decays and shake-off transitions taking place when xenon interacts with an 850 eV FEL pulse. To compute the single-photon ionisation cross-sections and Auger rates for xenon, we adapt to atoms  the formalism we have previously developed for diatomic molecules \cite{Banks2017}. This formalism is general and is applicable to complex multi-electron atoms  where the orbital wave functions are accurately described by many $l$ quantum numbers. Moreover, we have identified the prevalent pathways for the formation of each of the final ion states of xenon. As expected, comparing our results for pulses of the same duration and different peak intensity, we find that the highly-charged ion states have higher yields for the higher-intensity pulse. This is due to the larger number of single-photon
absorptions  taking place compared to the lower-intensity pulse. Also, we have identified the prevalent pathways for the ion states of xenon driven by two FEL pulses of the same pulse energy. We find that the longer-duration pulse has higher yields for the highly-charged ion states compared to the shorter-duration  pulse. This is due to the larger number of Auger transitions taking place between subsequent single-photon ionisation events in the case of the longer-duration pulse.

\section{Authors contributions}
All the authors were involved in the preparation of the manuscript.
All the authors have read and approved the final manuscript.
%

\bibliographystyle{IEEEtran}

\bibliography{Auger} 

\end{document}